\documentclass[aps,prd,preprint,superscriptaddress,amsfonts,amssymb,amsmath,nofootinbib]{revtex4-1}
\usepackage{graphicx,bm,color,ulem,latexsym}
\usepackage[colorlinks,linkcolor=magenta,anchorcolor=blue,citecolor=blue]{hyperref}
\usepackage{multirow}

\begin{document}
	
%%%%%%%%%%%%%%%%%%%%%%%%%%%%%%%%%%%%%%%%%%%%%%%%%%%%%%%%%%%%%%%%%%%%%%%%%%
\title{Solar system tests in covariant \texorpdfstring{$f(Q)$}{TEXT} gravity}
%%%%%%%%%%%%%%%%%%%%%%%%%%%%%%%%%%%%%%%%%%%%%%%%%%%%%%%%%%%%%%%%%%%%%%%%%%

\author{Wenyi Wang}
\email{wangwy@mails.ccnu.edu.cn}
\affiliation{Institute of Astrophysics, Central China Normal University, Wuhan 430079, China}
\author{Kun Hu}
\email{hukun@mails.ccnu.edu.cn (corresponding author)}
\affiliation{Institute of Astrophysics, Central China Normal University, Wuhan 430079, China}
\author{Taishi Katsuragawa}
\email{taishi@ccnu.edu.cn}
\affiliation{Institute of Astrophysics, Central China Normal University, Wuhan 430079, China}

\begin{abstract}

We study the Solar System constraints on covariant $f(Q)$ gravity. 
The covariant $f(Q)$ theory is described by the metric and affine connection, where both the torsion and curvature vanish.
Considering a model including a higher nonmetricity-scalar correction, $f(Q)= Q +\alpha Q^{n} - 2\Lambda$, 
we derive static and spherically symmetric solutions, which represent the Schwarzschild-de Sitter solution with higher-order corrections, for two different ansatz of the affine connection.  
On the obtained spacetime solutions, we investigate the perihelion precession, light deflection, Shapiro delay, Cassini constraint, and gravitational redshift in the $f(Q)$ gravity.
We place bounds on the parameter $\alpha$ with $n=2, 3$ in our model of $f(Q)$ gravity, using various observational data in the Solar System.

\end{abstract}

\maketitle	

%%%%%%%%%%%%%%%%%%%%%%%%%%%%%%%%%%%%%%%%%%%%%%%%%%%%%%%%%%%%%%%%%%%%%%%%%%
%%%%%%%%%%%%%%%%%%%%%%%%%%%%%%%%%%%%%%%%%%%%%%%%%%%%%%%%%%%%%%%%%%%%%%%%%%

\section{Introduction}

Einstein's theory of general relativity (GR) is the most successful theory of gravitation, 
however, it is plagued by several problems in a wide range of energy scales; for instance, the accelerated expansion of the current Universe~\cite{Capozziello:2011et} and the non-renormalizability in its canonical quantization~\cite{Stelle:1976gc}. 
Modified gravity can provide us with the possibility of explaining longstanding issues in GR.
The straightforward way of constructing such an extended theory of gravity is to add extra terms into the Einstein-Hilbert action.
One of the simple ways is to extend the Ricci scalar $R$ to its functional form $f(R)$~\cite{DeFelice:2010aj, Nojiri:2017ncd}.
However, an even more inventive approach would be to modify GR from a purely geometrical viewpoint.
Although GR describes gravity in terms of the spacetime curvature,
attempts to interpret gravity as a presence of nonmetricity $Q$ in spacetime instead of the curvature received attention recently.
This theory is known as symmetric teleparallel general relativity (STGR)~\footnote{
This theory is also called Symmetric Teleparallel Equivalent to General Relativity (STEGR).}~\cite{Adak:2005cd, Adak:2008gd, BeltranJimenez:2017tkd}.

In analogy to the way of extending GR to $f(R)$ gravity, STGR can be generalized to $f(Q)$ gravity characterized by the function of the nonmetricity scalar, called $f(Q)$ theory~\cite{BeltranJimenez:2017tkd, Heisenberg:2018vsk}.
It is already known that $f(Q)$ theory is not equivalent to $f(R)$ gravity, while STGR is equivalent to GR.
This theory has been vastly investigated in recent years, 
which has revealed interesting features and investigated various applications; for instance, cosmology~\cite{BeltranJimenez:2019tme, Bajardi:2020fxh, Anagnostopoulos:2021ydo, Dimakis:2022rkd, Lymperis:2022oyo, Shabani:2023xfn, Hu:2023ndc, Sokoliuk:2023ccw, Heisenberg:2023wgk, Sharif:2024arm, Mohanty:2024pzh, Wang:2024eai, Pradhan:2024eew}, 
and the physical degrees of freedom~\cite{Tomonari:2023wcs, DAmbrosio:2023asf, Heisenberg:2023lru, Hu:2022anq, Hu:2023gui, Gomes:2023tur}.
In particular, the spherically symmetric solutions in $f(Q)$ theory have been intensively studied~\cite{DAmbrosio:2021zpm, Zhao:2021zab, Wang:2021zaz, Hassan:2022hcb, Tayde:2022lxd, Parsaei:2022wnu, Kiroriwal:2023nul, Das:2024ytl, Dimakis:2024fan, Tayde:2024koy, Kumar:2024iaq}.

In the usual formulation of $f(Q)$ gravity, the general affine connection is assumed to be precisely eliminated $\Gamma^{\alpha}_{\ \mu\nu}= 0$ by employing diffeomorphism, 
and the choice of such an affine connection is known as the Coincident Gauge (CG). 
Although the CG is valid in solving the spherically symmetric solutions, as shown in Ref.~\cite{Zhao:2021zab, Wang:2021zaz}, 
it may lead to some undesirable features: 
the non-diagonal components of field equations require that $f(Q)$ is a linear function of $Q$ or $Q$ is a constant because the CG fails to be spherically symmetric~\cite{DAmbrosio:2021zpm}.
This undesirable feature can be cured by considering the covariant $f(Q)$ gravity with nonvanishing affine connection $\Gamma^{\alpha}_{\ \mu\nu} \neq 0$~\cite{Zhao:2021zab, Dimakis:2022rkd, Paliathanasis:2023nkb}. 

These distinct nonvanishing connections are obtained through the symmetry reduction method which satisfies the torsionless and curvatureless conditions. 
Particularly, Ref.~\cite{DAmbrosio:2021zpm} analyzed the weak-field solution obtained in $\Gamma^{\alpha}_{\ \mu\nu} \neq 0$ in a model $f(Q) = Q + \alpha Q^{n}$, where $\alpha$ is a constant and $n \ge 2$ is the integer.
$\alpha$ parametrizes the departure form GR or STGR, which provides the basis for constraining the $f(Q)$ theory using the spherically symmetric solution and the relevant observables in the Solar System.

This work aims to study the static and spherically symmetric solutions employed by two ansatz of the nonvanishing affine connection $\Gamma^{\alpha}_{\ \mu\nu}$ given by Ref.~\cite{DAmbrosio:2021zpm}.
Using a model $f(Q)= Q +\alpha Q^{n} - 2\Lambda$, we investigate the perihelion precession, light deflection, Shapiro delay, Cassini constraint, and gravitational redshift in the covariant $f(Q)$ gravity. 
Finally, we examine constraints on the model parameter $\alpha$ based on the observational data in two different choices of affine connection.

This paper is organized as follows. 
In Sec.~\ref{spherically symmetric Solutions}, we briefly review $f(Q)$ theory, where the action, the equations of motion, and spherically symmetric solutions are introduced.
In Sec.~\ref{solar system tests}, we formulate the perihelion precession, light deflection, Shapiro delay, Cassini constraint, and gravitational redshift in the $f(Q)$ gravity.
In Sec.~\ref{solar system constraints}, we use observational data in the Solar System to constrain the parameter $\alpha$. Finally, Sec.~\ref{conclusions} is devoted to the discussion.
Throughout the paper, we work in the geometrical unit that $c=G=1$.

%%%%%%%%%%%%%%%%%%%%%%%%%%%%%%%%%%%%%%%%%%%%%%%%%%%%%%%%%%%%%%%%%%%%%%%%%%
%%%%%%%%%%%%%%%%%%%%%%%%%%%%%%%%%%%%%%%%%%%%%%%%%%%%%%%%%%%%%%%%%%%%%%%%%%
	
\section{The Static and spherically symmetric Solutions}
\label{spherically symmetric Solutions}

This section briefly reviews the basics of the $f(Q)$ gravity, and we derive the two static and spherically symmetric solutions for two different affine connections. 
$f(Q)$ gravity is characterized by the following action~\cite{BeltranJimenez:2018vdo}
\begin{align}
\label{Eq: action}
    S 
    = 
    \frac{1}{16\pi} \int ~d^4x \sqrt{-g} f(Q) + S_{matter}
    \, .
\end{align}
$g$ is the determinant of the metric $g_{\mu\nu}$, 
and $f(Q)$ is an arbitrary function of the nonmetricity scalar $Q$. 
Nonmetricity tensor $Q_{\alpha \mu \nu}$ is defined by the metric and affine connection $\Gamma^{\alpha}_{\ \mu\nu}$,
\begin{align}
\begin{split}
    Q_{\alpha \mu \nu}
    &\equiv 
    \nabla _{\alpha }g_{\mu \nu}
    \\
    &= 
    \partial_{\alpha}g_{\mu \nu} 
    - \Gamma^{\lambda}_{\ \alpha\mu}g_{\lambda \nu} 
    - \Gamma^{\lambda}_{\ \alpha\nu}g_{\mu \lambda}
    \, .
\end{split}
\end{align}
Nonmetricity scalar $Q$ is defined as 
\begin{align}
    Q
    =
    -Q_{\alpha \mu \nu }P^{\alpha \mu \nu}
    \, ,
\end{align} 
where $P^{\alpha }_{\ \mu \nu }$ is called the nonmetricity conjugate,
\begin{align}
\label{Eq: nms}
    P^{\alpha }_{\ \mu \nu } 
    &= 
    -\frac{1}{4}Q^{\alpha}_{\ \mu \nu } +\frac{1}{2}Q_{(\mu \ \nu )}^{\ \  \alpha} 
    +\frac{1}{4}\left( Q^{\alpha } -\tilde{Q}^{\alpha }\right)g_{\mu \nu }
    -\frac{1}{4}\delta^{\alpha}_{\ (\mu} Q_{\nu)}
    \, .
\end{align}
$Q_{\alpha}\equiv Q_{\alpha \ \mu}^{\ \mu}$ and $ \tilde{Q}_{\alpha } \equiv Q^{\mu}_{\ \alpha \mu }$ are traces of the nonmetricity tensor.

Varying Eq.~\eqref{Eq: action} with respect to the metric and connection, respectively, we obtain the field equations of $f(Q)$ gravity:
\begin{align}
\label{Eq: metric-eom}
    0
    &= 
    \frac{2}{\sqrt{-g}}\nabla_{\alpha} \left(\sqrt{-g} f_{Q} P^{\alpha}_{\ \mu\nu} \right)
    + \frac{1}{2}g_{\mu\nu} f  
    + f_Q \left( P_{\mu\alpha\beta}Q_{\nu}^{\ \alpha\beta}
    -2 Q_{\alpha\beta\mu}P^{\alpha\beta}_{\ \ \ \nu}\right) + 8\pi T_{\mu\nu} 
    \, ,
    \\
    0
    &= 
    \nabla_\mu\nabla_\nu \left(\sqrt{-g} f_Q P^{\mu\nu}_{\ \ \alpha} \right) 
    \, ,
\end{align}
where we denote the derivative of $f(Q)$ with respect to $Q$ by $f_Q \equiv \partial_{Q}f(Q)$.
Ref.~\cite{DAmbrosio:2021zpm} has shown that using the symmetry reduction of metric and connection imposed by the torsionless and stationary condition, we obtain the following expressions in the spherical coordinate system $(t, r, \theta, \phi)$:
\begin{align}
\label{field-equation-gtt}
\begin{split}
    \partial _{r }g_{tt} 
    &=
    - \frac{g_{tt} \left[2 f_{Q} +g_{rr}\left(Q r^{2} f_{Q} -2f_{Q}-f r^{2}\right)\right]}{2 r f_{Q}}
    % \\
    % & \qquad \qquad
    + \frac{g_{tt}\left[r^{2} - g_{rr} (\Gamma^{r}_{\ \theta \theta})^{2}\right]}{\Gamma^{r}_{\ \theta \theta} r f_{Q}} 
    f_{QQ} \partial _{r}Q
    \, ,
\end{split}
    \\
\label{field-equation-grr}
\begin{split}
    \partial _{r }g_{rr} 
    &= 
    \frac{g_{rr} \left[2 f_{Q} + g_{rr}\left(Q r^{2} f_{Q} -2f_{Q}-f r^{2}\right)\right]}{2 r f_{Q}}
    % \\
    % & \qquad \qquad
    + \frac{g_{rr}\left[r^{2} + g_{rr} (\Gamma^{r}_{\ \theta \theta})^{2} +2 r \Gamma^{r}_{\ \theta \theta} \right]}{\Gamma^{r}_{\ \theta \theta} r f_{Q}}
    f_{QQ} \partial _{r}Q
    \, ,
\end{split}
    \\
\label{nonmetricity-scalar}
\begin{split}
    Q 
    &= 
    -\frac{1}{r^{2} (\Gamma^{r}_{\ \theta \theta})^{2} g_{tt} g_{rr}^{2}}
    \\
    & \qquad \times
    \Big \{ 
        \partial_{r }g_{tt} \Gamma^{r}_{\ \theta \theta} g_{rr}\left[ g_{rr}(\Gamma^{r}_{\ \theta \theta})^{2} + r (r + 2\Gamma^{r}_{\ \theta \theta}) \right]
    \\
    & \qquad \qquad
        + g_{tt} \left[-2 g_{rr}^{2} (\Gamma^{r}_{\ \theta \theta})^{3} \Gamma^{r}_{\ r r} -r^{2} \Gamma^{r}_{\ \theta \theta} \partial_{r} g_{rr}\right] 
    \\
    & \qquad \qquad
        + g_{tt} g_{rr} \left[2 r^{2} + 4 r \Gamma^{r}_{\ \theta \theta} + 2 (\Gamma^{r}_{\ \theta \theta})^{2} 
        + 2 r^{2} \Gamma^{r}_{\ \theta \theta} \Gamma^{r}_{\ r r} + (\Gamma^{r}_{\ \theta \theta})^{3} \partial_{r}g_{rr}\right]
    \Big \}
    \, ,
\end{split}
\end{align}
where $f_{QQ} \equiv \partial_{QQ}f(Q)$.

The only nonvanishing component of affine connection in Eqs.~\eqref{field-equation-gtt}-\eqref{nonmetricity-scalar} is $\Gamma^{r}_{\ \theta \theta}$, which satisfies the following differential relation,
\begin{align}
\label{diffeq-Gamma-r-theta-theta}
    \partial_{r} \Gamma^{r}_{\ \theta \theta} 
    = 
    -1 -\Gamma^{r}_{\ \theta \theta} \Gamma^{r}_{\ r r}
    \, .
\end{align}
Regarding the metric components, the line element for a generic static and spherically symmetric spacetime can be written as
\begin{align}
\label{Eq: spherical-ansatz}
    ds^{2}
    =
    -A(r)dt^{2}+B(r)dr^{2} + r^{2}d\Omega^{2}
    \, ,
\end{align}
where the $A(r) = -g_{tt}$ and $B(r) = g_{rr}$ and $d\Omega ^2=d\theta ^2+\sin ^2\theta d\varphi ^2$.
The nonvanishing component of affine connection the $\Gamma^{r}_{\ \theta\theta}$ can be choose as (more details see Ref.\cite{DAmbrosio:2021zpm})
\begin{align}
\label{Gamma-r-theta-theta}
    \Gamma^{r}_{\ \theta\theta}
    = 
    \pm \frac{r}{\sqrt{B(r)}}
    \, .
\end{align}

In this work, we consider a Lagrangian in the form of $f(Q)= Q + \alpha Q^{n} - 2\Lambda$ for $n=2,3$. 
The $\Lambda$ represents the cosmological constant, and $\alpha$ is a small constant in our model.
Using the perturbative treatment of the high-order correction of the nonmetricity, we can suppose that the correction to GR or STGR will lead to deviations characterized by $\alpha$ from the ordinary Schwarzschild-de Sitter solutions:
\begin{align}
    A(r) 
    &= 
    g_{tt}^{(0)} + \alpha g_{tt}^{(1)}
    \, ,
    \\
    B(r) 
    &= 
    g_{rr}^{(0)} + \alpha g_{rr}^{(1)}
    \, .
\end{align}
$g_{tt}^{0}$ and $g_{rr}^{0}$ are ‌corresponding to components of the metric of the Schwarzschild-de Sitter solution in weak-gravity limit
\begin{align}
    g_{tt}^{(0)} 
    = 
    1 - \frac{2M}{r} - \frac{1}{3} \Lambda r^{2}
    \, ,
    \\
    g_{rr}^{(0)} 
    = 
    1 + \frac{2M}{r} + \frac{1}{3} \Lambda r^{2}
    \, .
\end{align}
Employing Eq.~\eqref{Gamma-r-theta-theta}, we solve Eqs.~\eqref{field-equation-gtt}-\eqref{diffeq-Gamma-r-theta-theta} and derive the metric components $A(r)$ and $B(r)$. Hereafter, the case $\Gamma^{r}_{\ \theta\theta}= - r/ \sqrt{B(r)}$ will be referred to as Case I, and another $\Gamma^{r}_{\ \theta\theta}= + r/ \sqrt{B(r)}$ as Case II.

In Case I, we obtain the following solutions:
\begin{align}
\label{caseI-A}
    A(r) 
    &= 
    1 - \frac{2M}{r} - \frac{1}{3}\Lambda r^{2}
    - \alpha \cdot \frac{(-1)^{n} 2^{n} n M^{2n-1}}{(4n-3)r^{4n-3}}
    \, , 
    \\
\label{caseI-B}
    B(r) 
    &= 
    1 + \frac{2M}{r} + \frac{1}{3}\Lambda r^{2}
    + \alpha \cdot \frac{(-1)^{n} 2^{n} n M^{2n-1}}{r^{4n-3}}
    \, ,
\end{align}	
and the nonmetricity scalar is given by
\begin{align}
\label{nonmetricity scalar-caseI}
    Q 
    = 
    \frac{2 M^{2}}{r^{4}} 
    + \alpha \cdot \frac{(-1)^{n} 2^{n+1} n M^{2}}{r^{4n}}
    \, .
\end{align}
In Case II, 
\begin{align}
\label{caseII-A}
    A(r) 
    &= 
    1-\frac{2M}{r}  - \frac{1}{3}\Lambda r^{2}
    - \alpha \cdot \frac{2^{3n-1}}{(2n-3)r^{2n-2}}  
    \, ,
    \\
\label{caseII-B}
    B(r) 
    &= 
    1 + \frac{2M}{r} +\frac{1}{3}\Lambda r^{2}
    + \alpha \cdot \frac{2^{3n-1}\left(2n^{2}-3n+1 \right)}{(2n-3)r^{2n-2}}
    \, .
\end{align}	
and 
\begin{align}
\label{nonmetricity scalar-caseII}
    Q 
    = 
    \frac{8}{r^{2}} - \alpha \cdot \frac{2^{3n+1} (n-1)}{r^{2n}}
    \, .
\end{align}

%%%%%%%%%%%%%%%%%%%%%%%%%%%%%%%%%%%%%%%%%%%%%%%%%%%%%%%%%%%%%%%%%%%%%%%%%%
%%%%%%%%%%%%%%%%%%%%%%%%%%%%%%%%%%%%%%%%%%%%%%%%%%%%%%%%%%%%%%%%%%%%%%%%%%

\section{Calculations for Solar System tests}
\label{solar system tests}

In this section,
we investigate the Solar-System tests, including the perihelion precession, light deflection, Shapiro delay, Cassini constraint, and gravitational redshift, in $f(Q)$ gravity.
We follow methodologies developed in Ref.~\cite{Farrugia:2016xcw}.
Although Ref.~\cite{Farrugia:2016xcw} discusses the Solar-System constraint in $f(T)$ gravity,
which is another type of geometrical modification of gravitational theory,
the background spacetime solution therein corresponds to our Case II solution.
Thus, comparing Cases I and II allows us to discuss the impacts of different choices of affine connections.
After we establish the general formulation in static and spherically symmetric background as in Eq.~\eqref{Eq: spherical-ansatz}, we apply two sets of the metric and affine connection obtained in the previous section.

%%%%%%%%%%%%%%%%%%%%%%%%%%%%%%%%%%%%%%%%%%%%%%%%%%%%%%%%%%%%%%%%%%%%%%%%%%

\subsection{Perihelion Precession}
\label{sec:Perihelion Precession}

We investigate the perihelion precession.
We begin with the general form of the static and spherically symmetric solution in $f(Q)$ gravity written as follows:
\begin{align}
\label{general case}
\begin{split}
    A(r) 
    &= 
    1-a_{1}(r)\frac{M}{r}+a_{2}(r)\frac{M^{2}}{r^{2}}
    \, ,
    \\ 
    B(r) 
    &= 
    1+b_{1}(r)\frac{M}{r}+b_{2}(r)\frac{M^{2}}{r^{2}}
    \, .
\end{split}
\end{align}	
For the metric given by Eq.~\eqref{Eq: spherical-ansatz}, 
the normalization condition for four-velocity $u^{\mu}$, $g_{\mu \nu} u^{\mu} u^{\nu} = -\eta$, can be written as
\begin{align}
\label{four-velocity normalization}
    \eta
    = 
    A(r) \: \left(\frac{dt}{d\lambda}\right)^2 
    - B(r) \: \left(\frac{dr}{d\lambda}\right)^2 
    - r^2 \: \left(\frac{d\phi}{d\lambda}\right)^2 
    \, .
\end{align}
where $\lambda$ is the affine parameter. 
For a timeline ($\eta=1$) test particle, Eq.~\eqref{four-velocity normalization} gives 
\begin{align}
\label{timelike-metric}
    1
    = 
    A(r) \: \left(\frac{dt}{d\tau}\right)^2 
    - B(r) \: \left(\frac{dr}{d\tau}\right)^2 
    - r^2 \: \left(\frac{d\phi}{d\tau}\right)^2 
 \, .
\end{align}

For the static and spherically symmetric spacetime~\eqref{Eq: spherical-ansatz}, 
the metric is invariant under time translation and $SO(3)$ spatial rotation, 
which gives two kinds of Killing vectors.
One is the timelike Killing vector $\mathring{\xi}^{\mu}=\left( 1,0,0,0 \right) $, and another is the space rotational Killing vector $\bar{\xi}_{\mu}=\left( 0,0,0,r^2 \sin ^2\theta \right) $. 
Considering the equatorial movement of the test particle, i.e. $\theta=\pi/2$, 
we can define two kinds of conserved quantities 
by using the fact that $\xi _{\mu}u^{\mu}=\text{const.}$ along the geodesics:
\begin{align}
\label{energy-momentum-timelike} 
    E 
    = 
    A(r)\frac{dt}{d\tau}
    \, , \quad
    L 
    = 
    r^{2} \frac{d\phi}{d\tau}
    \, .
\end{align}
$E$ and $L$ are interpreted as the total energy and angular momentum per unit mass of the particle measured by the observer from infinity, respectively. 
Upon the use of the above formula, we can rewrite Eq.~\eqref{timelike-metric} in terms of the conserved quantities as follows:
\begin{align}
\label{perihelion}
    1 
    = 
    \frac{E^{2}}{A(r)} 
    - \frac{B(r) L^{2}}{r^{4}} \left(\frac{dr}{d\phi}\right)^{2} 
    - \frac{L^{2}}{r^{2}}
    \, .
\end{align}

Using Eq.~\eqref{perihelion}, we find the angular distance between the perihelion $r_{-}$ and aphelion $r_{+}$,
\begin{align}
\label{integral-perihelion}
\begin{split}
    \phi(r_+) - \phi(r_-) 
    &=
    \int_{\phi(r_{-})}^{\phi(r_{+})} d\phi 
    \\
    &= 
    \int_{r_{-}}^{r_{+}} 
    \frac{B(r)^{1/2}}{r^{2}} 
    \left(
        \frac{E^{2}}{L^{2} A(r)} - \frac{1}{L^{2}} - \frac{1}{r^{2}} 
    \right)^{-1/2} \: dr
    \, .
\end{split} 
\end{align}
This is followed by the total perihelion precession for per orbit
\begin{align}\label{total-perihelion-angle}
    \Delta \phi 
    = 
    2|\phi(r_+)-\phi(r_-)| - 2\pi
    \, .
\end{align}
Using the fact that $dr/d\phi$ vanishes at $r_{-}$ and $r_{+}$, which puts the Eq.~\eqref{perihelion} into the following form
\begin{align}
    1
    =
    \frac{E^{2}}{A(r_{\pm})}-\frac{L^{2}}{r_{\pm}^{2}} 
    \, ,
\end{align}
we can derive the following values for the $E$ and $L$:
\begin{align}
\label{E-value}
    E^2 
    &= 
    \frac{A(r_-) A(r_+) \left(r_{+}^2 - r_{-}^2\right)}
    {A(r_{-})r_{+}^{2} - A(r_{+})r_{-}^{2}}
    \, , 
    \\
\label{L-value}
    L^2 
    &= 
    \frac{r_{+}^{2} r_{-}^{2} \left( A(r_{-})-A(r_{+})\right)}
    {A(r_{+}){r_{-}}^{2}-A(r_{-}){r_{+}}^{2}} 
    \, ,
\end{align}

Plugging the Eqs.~\eqref{E-value}~\eqref{L-value} into Eq.~\eqref{integral-perihelion}, we obtain the following expression:
\begin{align}
\label{integral-perihelion-1}
\begin{split}
    &\phi(r_+) - \phi(r_-) 
    \\
    &= 
    \int_{r_-}^{r_+}dr\; 
    \frac{B(r)^{1/2}}{r^2} 
    \left \{
        \frac{A(r_-) A(r_+) \left(r_{+}^2 - r_{-}^2\right) + A(r) \left[ A(r_+)r_{-}^2-A(r_-) r_{+}^2\right]}
        {r_{+}^2 r_{-}^2 \left[ A(r_+)-A(r_-)\right] A(r)} 
        - \frac{1}{r^2} 
    \right \}^{-1/2} 
    \, . 
\end{split}
\end{align}
Note that the bracketed part in Eq.~\eqref{integral-perihelion-1} vanishes at $r = r_{\pm}$. 
Thus, we can rewrite the second square root in Eq.~\eqref{integral-perihelion} as a quadratic function of $1/r$:
\begin{align}
\label{C-equation}
\begin{split}
    & \frac{A(r_-) A(r_+) \left(r_{+}^2 - r_{-}^2\right) + A(r)\left(A(r_+)r_{-}^2-A(r_-) r_{+}^2\right)}{r_{+}^2 r_{-}^2 \left(A(r_+)-A(r_-)\right) A(r)} - \frac{1}{r^2} 
    \\
    & = 
    C(r) \left(\frac{1}{r} - \frac{1}{r_+}\right) \left(\frac{1}{r_-} - \frac{1}{r}\right)
    \, .
\end{split}
\end{align}
For the case of slightly elliptic orbits, we have $C(r)\simeq C = const.$. 
To determine the constant $C$, it is convenient to introduce the new variable $u =1/r$, which also can be written as:
\begin{align}
\label{semi-latus-def}
    u 
    \simeq 
    \frac{1}{2} \left(\frac{1}{r_{+}} + \frac{1}{r_{-}} \right) 
    = 
    \frac{1}{a (1 - e^{2})}
    \, ,
\end{align}
where $a$ is the semimajor axis and $e$ is the orbital eccentricity. 
Differentiating both sides of the Eq.~\eqref{C-equation} yields:
\begin{align}
\label{C-equation-1}
    C 
    = 
    1 - \frac{(u_+ - u_-)(u_+ + u_-)}{2 \left[ X(u_+) - X(u_-)\right]} X^{\prime \prime}(u)
    \, , 
\end{align}
where $X(u)$ is defined by $X(u)=A(u)^{-1}$, we also have
\begin{align}
	X(u_+) - X(u_-) \approx (u_+ - u_-) X'(u)
    \, ,
\end{align}
which simplifies Eq.~\eqref{C-equation-1} to
\begin{align}
\label{C-equation-2}
    C 
    = 
    1 - \frac{u X^{\prime \prime}(u)}{X^{\prime}(u)}
    \, .
\end{align}
With the help of Eq.~\eqref{C-equation}, the expression of angular distance \eqref{integral-perihelion-1} can be simplified as
\begin{align}
\label{integral-perihelion-modified-1}
    \phi(r_+) - \phi(r_-) 
    \simeq  
    \int_{r_-}^{r_+} \frac{B(r)^{1/2}}{r^2} 
    \left[
        C \left(\frac{1}{r} - \frac{1}{r_+}\right) 
        \left(\frac{1}{r_-} - \frac{1}{r}\right)
    \right]^{-1/2} \: dr
    \, .
\end{align}

Next, we perform the following change of the variable,
\begin{align}
    u 
    = 
    \frac{1}{2} (u_+ + u_-) 
    + \frac{1}{2} (u_+ - u_-) \sin \psi
    \, ,
\end{align}
while $\psi$ is a new integral variable. 
At the point of perihelion and aphelion, the corresponding values of $\psi$ are taken to be $\psi=-\pi/2$ and $\psi=\pi/2$. 
Therefore, Eq.~\eqref{integral-perihelion-modified-1} can be written in terms of the new angular variable $\psi$:
\begin{align}
\label{integral-perihelion-modified-2}
    \phi(r_+) - \phi(r_-) 
    = 
    C^{-1/2} \int_{-\pi/2}^{\pi/2} B(\psi)^{1/2} \: d\psi
    \, .
\end{align}	
Utilizing the Eqs.~\eqref{general case} and \eqref{C-equation-2}, 
we expand the constant $C$ up to the first order in $a_{1}(u)$ and $a_{2}(u)$
% and neglecting their products
, which is found to be:
\begin{align}
\label{expand C}
\begin{split}
    C 
    & \simeq  
    1 - 2  M u a_{1}(u) - 2 M u^{2} a_{1}^{\prime}(u) + 2 M^{2} u^{3} a_{2}^{\prime}(u) + 4 M^{2} u^{2} a_{2}(u) 
    \\
    & \qquad 
    + \frac{1}{a_{1}(u)}
    \left[
        2 M u a_{2}(u) - 2 u a_{1}^{\prime}(u) + 4 M u^{2} a_{2}^{\prime}(u) 
        - u^{2} a_{1}^{\prime\prime}(u) + M u^{3} a_{2}^{\prime\prime}(u)
    \right]
    \, .
\end{split}
\end{align}
Similarly, we carry out the integral in Eq.~\eqref{integral-perihelion-modified-2} up to first order in  $b_{1}(u)$ and $b_{2}(u)$, and neglecting their products
\begin{align}
\label{expand B(psi)}
    \int_{-\pi/2}^{\pi/2} B(\psi)^{1/2} \: d\psi 
    \simeq
    \pi +\frac{1}{2} M u \pi b_{1}(u) + \frac{1}{2} M^{2}u^{2} \pi b_{2}(u)
    \, .
\end{align}
Putting Eqs.~\eqref{expand B(psi)} and \eqref{expand C} together and inserting them into Eqs.~\eqref{total-perihelion-angle} and \eqref{integral-perihelion-modified-2}, we obtain the total perihelion precession for the case of slightly elliptic orbits:
\begin{align}
\label{perihelion-general-solution}
\begin{split}
    \Delta\phi 
    &\approx 
    \frac{1}{2} \pi  u 
    \Big \{ 
        4 M a_{1}(u) 
        + 2 M b_{1}(u) 
        - 8 M^{2} u a_{2}(u) 
    \\
    & \qquad \qquad
        + 2 M u 
        \left[ 
            M b_{2}(u) + 2a_{1}^{\prime}(u) - 2 M u a_{2}^{\prime}(u) 
        \right]
    \\
    & \qquad \qquad
        - \frac{1}{a_{1}(u)} 
        \left[ 2 + M u b_{1}(u) + M^{2}u^{2}b_{2}(u) \right]
    \\
    & \qquad \qquad \quad
        \times
        \left[ 
            2 M a_{2}(u) 
            - 2a_{1}^{\prime}(u) 
            + 4 M u a_{2}^{\prime}(u) 
            - u a_{1}^{\prime\prime}(u) 
            + M u^{2} a_{2}^{\prime\prime} 
        \right]
   \Big \}
   \, .
\end{split}
\end{align}

To illustrate our results, we now apply the general expression Eq.~\eqref{perihelion-general-solution} to the specific solutions we demonstrated in Sec II.
Reading $a_{1}(u)$, $a_{2}(u)$, $b_{1}(u)$ and $b_{2}(u)$ from results in Case I and II,
we formulate the perihelion precessions in these two cases.
For Case I, we find
\begin{align}
\begin{split}
    a_{1}(u) 
    &=
    2 
    \, , \quad 
    a_{2}(u) 
    =
    - \frac{\Lambda}{3 M^{2}}u^{-4} 
    - \alpha  \cdot \frac{(-1)^{n} 2^{n} n}{4n-3} \cdot M^{2n-3} u^{4n-5}
    \, ,
    \\
    b_{1}(u) 
    &=
    2 
    \, , \quad 
    b_{2}(u)
    =
    \frac{\Lambda}{3 M^{2}}u^{-4}
    + 
    \alpha \cdot (-1)^{n} 2^{n} n \cdot M^{2n-3} u^{4n-5}
    \, .
\end{split}
\end{align}
Substituting $a_{1}^{\prime}(u), b_{1}^{\prime}(u), a_{1}^{\prime\prime}(u), b_{1}^{\prime\prime}(u)$ into Eq.~\eqref{perihelion-general-solution}, 
we obtain the total perihelion precession in Case I:
\begin{align}
\label{case I-solution}
\begin{split}
    \Delta\phi 
    & \approx 
    \frac{6 \pi  M}{a(1-e^2)}+\frac{\pi  \Lambda a^3(1-e^2)^3}{M}
    \\
    & \qquad
    +
    \alpha \cdot \pi \left[ (-2)^{n+1}+n(-1)^{n}2^{n+1} \right] n \cdot 
    \frac{M^{2n-2}}{a^{4n-4} (1-e^2)^{4n-4}} 
    \\
    & \qquad
    + 
    \alpha \cdot \frac{\pi [3(-2)^{n}+(-2)^{n+1} n+(-1)^{n}n^{2}2^{n+3}] n}{4 n-3} \cdot
    \frac{M^{2n-1}}{a^{4n-3} (1-e^2)^{4n-3}} 
    \, ,
\end{split}
\end{align}
When we choose $n=2,3$, the following solutions are obtained respectively:
\begin{align}
\label{perihelion-solution-I2}
\begin{split}
    \Delta \phi^{\text{I}}_{\text{n=2}}
    &\approx 
    \frac{6 \pi  M}{a(1-e^{2})}
    + \frac{\pi  \Lambda a^{3}(1-e^2)^{3}}{M}
    \\
    & \qquad
    + \alpha
    \left[
        \frac{16 \pi M^{2}}{a^{4} (1-e^2)^{4}}
        + \frac{40 \pi M^{3}}{a^{5} (1-e^2)^{5}}
    \right]
    \, , 
\end{split}
    \\
\label{perihelion-solution-I3}
\begin{split}
    \Delta \phi^{\text{I}}_{\text{n=3}} 
    &\approx 
    \frac{6 \pi  M}{a(1-e^{2})}
    + \frac{\pi  \Lambda a^{3}(1-e^2)^{3}}{M}
    \\
    & \qquad
    -  \alpha
    \left[
        \frac{96 \pi M^{4}}{a^{8} (1-e^2)^{8}}
        + \frac{168 \pi M^{5}}{a^{9} (1-e^2)^{9}}
    \right]
    \, .
\end{split} 
\end{align}

Similarly, for Case II, we find
\begin{align}
\begin{split}
    a_{1}(u) 
    &= 2 
    \, ,\quad 
    a_{2}(u) 
    = 
    - \frac{\Lambda}{3 M^{2}}u^{-4} 
    - \alpha \cdot \frac{2^{3n-1}}{2n-3} \cdot  \frac{u^{2n-4}}{M^{2}}
    \, ,
    \\
    b_{1}(u) 
    &=
    2 
    \, ,\quad 
    b_{2}(u)
    =
    \frac{\Lambda}{3 M^{2}}u^{-4}
    +
    \alpha \cdot \frac{2n^{2}-3n+1}{2n-3} \cdot \frac{u^{2n-4}}{M^{2}}
    \, .
\end{split}		
\end{align}
Eq.~\eqref{perihelion-general-solution} leads to the total perihelion precession in Case II:
\begin{align}
\label{case II-solution}
\begin{split}
    \Delta \phi 
    &\approx 
    \frac{6 \pi  M}{a(1-e^2)} 
    + \frac{\pi  \Lambda a^3(1-e^2)^3}{M}
    \\
    & \qquad
    + 
    \alpha \cdot \pi 2^{3 n-1} (n-1) \cdot \frac{1}{M a^{2n-3} (1-e^2)^{2n-3}} 
    \\
    & \qquad
    +
    \alpha \cdot \frac{\pi 2^{3 n+1} n(n-1) }{2n-3} \cdot \frac{1}{a^{2n-2} (1-e^2)^{2n-2}}
    \, .
\end{split}
\end{align} 
We find consistent results with Ref.~\cite{Farrugia:2016xcw} when we choose $n=2,3$ for the case II:
\begin{align}
\label{perihelion-solution-II2}
\begin{split}
    \Delta \phi^{\text{II}}_{\text{n=2}} 
    &\approx 
    \frac{6 \pi  M}{a(1-e^{2})} 
    + \frac{\pi  \Lambda a^{3}(1-e^2)^{3}}{M} 
    \\
    & \qquad
    + 
    \alpha 
    \left[
        \frac{32 \pi}{a \left(1-e^{2}\right) M} 
        + \frac{256 \pi}{a^{2}\left(1-e^{2}\right)^{2}}
    \right]
    \, ,
\end{split}
    \\
\label{perihelion-solution-II3}
\begin{split}
    \Delta \phi^{\text{II}}_{\text{n=3}} 
    &\approx 
    \frac{6 \pi  M}{a(1-e^{2})} 
    + \frac{\pi  \Lambda a^{3}(1-e^2)^{3}}{M}
    \\
    & \qquad
    + 
    \alpha
    \left[
    \frac{512 \pi}{a^{3}\left(1-e^{2}\right)^{3} M}
    + \frac{2048 \pi}{a^{4} \left(1-e^{2}\right)^{4}}
    \right]
    \, .
\end{split}
\end{align}

We can find the different dependence on source mass $M$ in Case I and Case II.
For Case I, the influence from the correction $Q^n$ in Eqs.~\eqref{perihelion-solution-I2}~\eqref{perihelion-solution-I3} is proportional to $M^2$ and $M^3$,
and thus, the deviation from GR grows with increasing $M$. 
In contrast, for Case II, the third term in Eq.~\eqref{perihelion-solution-II2}~\eqref{perihelion-solution-II3} is proportional to the $M^{-1}$, and the last term is independent of $M$.
These results imply that the difference of the perihelion precession between our model and GR becomes larger when $M$ is smaller in Case II.

%%%%%%%%%%%%%%%%%%%%%%%%%%%%%%%%%%%%%%%%%%%%%%%%%%%%%%%%%%%%%%%%%%%%%%%%%%

\subsection{Light deflection}
\label{sec:light-bending}

Next, applying the same setup as in the previous subsection,
we study the light deflection in the background given by the spherically symmetric spacetime~\eqref{general case}.
We focus on the corrections to the GR prediction of the deflection angle due to the modification of the Lagrangian.
For photon, we choose $\eta = 0$ in Eq.~\eqref{four-velocity normalization}, which yields
\begin{align}
\label{light-bending-1st-order-DE}
    0
    = 
    \frac{E^{2}}{L^{2}} 
    - \frac{A(r) B(r)}{r^4} \left(\frac{dr}{d\phi}\right)^{2} 
    - \frac{A(r)}{r^{2}}
    \, .
\end{align}
Similar to the previous section, $E$ and $L$ can be regarded as the energy and angular momentum of the photon, respectively~\footnote{Since the choice of the affine parameter $\lambda$ is arbitrary, the physical meaning of these two conserved quantities is ambiguous.} 
\begin{align}
\label{energy-momentum-null} 
    E 
    = 
    A(r) \frac{dt}{d\lambda}
    \, , \quad
    % \\
% \label{angular-momentum-null}
    L 
    = 
    r^2 \frac{d\phi}{d\lambda}
    \, .
\end{align}

To avoid their dependence on the affine parameter $\lambda$, we introduce the impact parameter $b$ defined by
\begin{align}
\label{impact parameter}
    b^{2} = \frac{L^{2}}{E^{2}}
    \, .
\end{align}
Differentiating Eq.~\eqref{light-bending-1st-order-DE} with respect to the new variable $u = 1/r$ and utilize Eq.~\eqref{impact parameter}, we find the following equation,
\begin{align}
\label{light-bending-2nd-order-DE}
    0
    = 
    \frac{d^2 u}{d\phi^2} + \frac{b^{-2} 
    - u^2 A(u)}{2 A(u)^2 B(u)^2} \frac{d}{du} \left[A(u)B(u) \right] 
    + \frac{1}{2 A(u)B(u)} \frac{d}{du} \left[u^2 A(u)\right]
    \, .
\end{align}
We now consider the general forms of metric components presented as $A(u)=1+A_0(u)$, $B(u)=1+B_0(u)$. 
Up to the first order of $A_0(u)$ and $B_0(u)$, the differential equation~\eqref{light-bending-2nd-order-DE} takes the following approximated form:
\begin{align}
\label{differential-equation}
    0
    \approx 
    \frac{d^2 u}{d\phi^2} + u\left(1 -B_0(u)\right) 
    + \frac{1}{2} {b}^{-2}  \frac{d}{du} \left[A_0(u) + B_0(u) \right] 
    - \frac{1}{2} u^2 \frac{d}{du} \left[B_0(u)\right]
    \, ,
\end{align}

\begin{figure}[htbp]
    \centering
    \includegraphics[height=6cm,width=12cm]{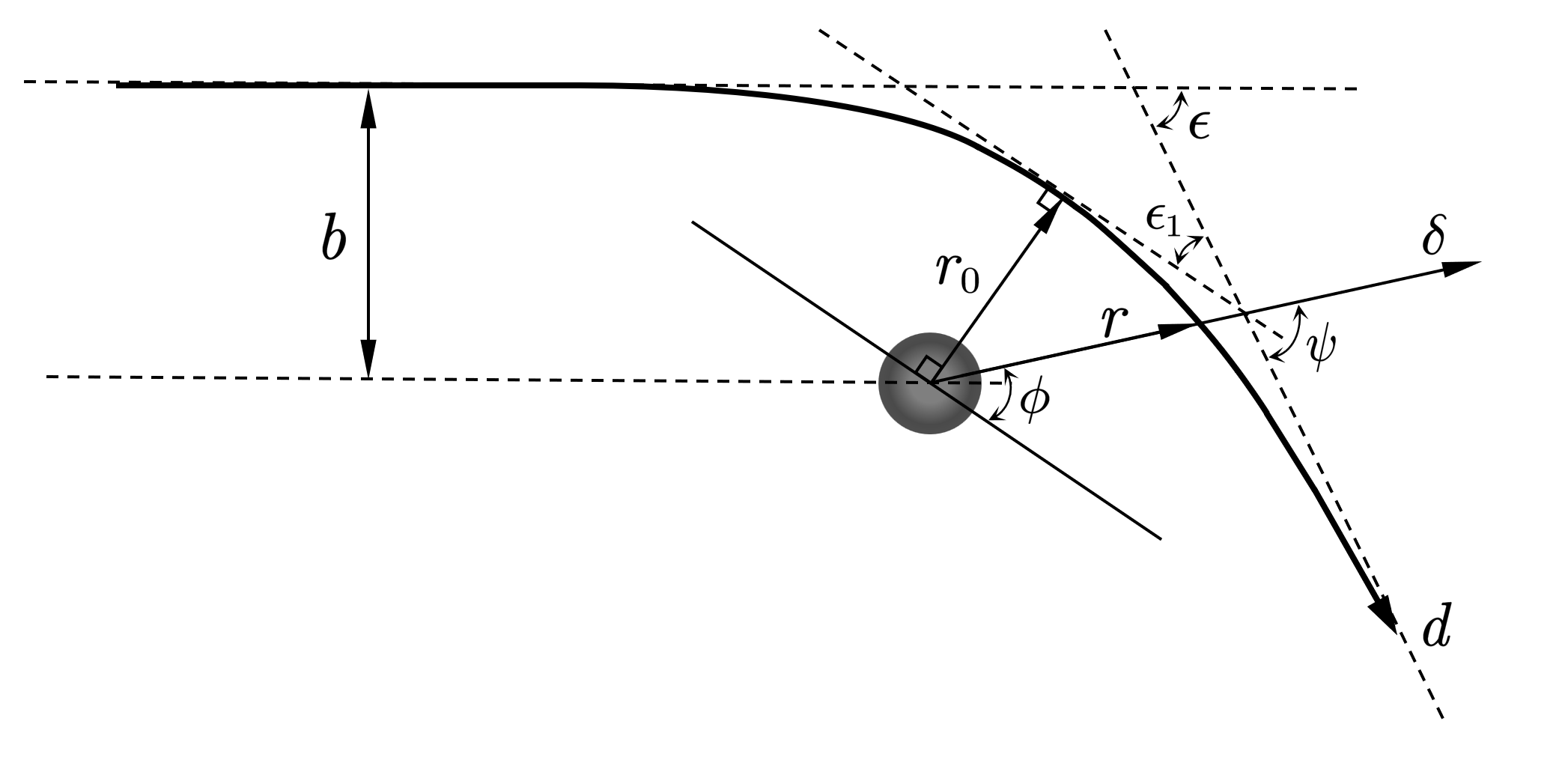}
    \caption{
    The plane graph represents the orbit with the one-sided deflection angle $\epsilon_1= \psi - \phi=\epsilon/2$. (we greatly exaggerate the deflection angle.) 
    }
    \label{deflection angle}
\end{figure}

To calculate the deflation angle, we adopt Rindler and Ishak's invariant cosine technique~\cite{Rindler:2007zz}, which in turn allows us to determine the one-sided deflection angle $\epsilon_{1}$ as shown in Fig.~\ref{deflection angle}. 

Considering two coordinate directions $\vec{d} = (dr,\: d\phi) $ and $\vec{\delta}= (\delta r,\: 0)$ with $\vec{d}$ represent the exiting direction of the light and $\vec{\delta}$ oriented toward the direction of the axis with $\phi=\text{const.}$, we have the following relation for the cosine of the angle $\psi$ between two coordinate directions $\vec{d}$ and $\vec{\delta}$:
\begin{equation}
\begin{aligned}
\label{invariant-cosine}
    \cos \psi 
    &= 
    \frac{<\bm{d}, \bm{\delta}>}
    {\sqrt{||\bm{d}||} \sqrt{||\bm{\delta} ||}}
    \\
    &= 
    \frac{g_{ij} d^i \delta^j}
    {\sqrt{g_{ij} d^i d^j} \sqrt{g_{ij} \delta^i \delta^j}}
    \, .
\end{aligned}
\end{equation}
The relevant $g_{ij}$ is
\begin{align}
	g_{11}=B(r)  	\, ,\quad    g_{22}=r^{2}
\end{align}
and the two coordinate directions 
\begin{align}
    \vec{d} 
    &= 
    (dr,\: d\phi) 
    =
    (D,\: 1) d\phi
    \, ,
    \\
    \vec{\delta} 
    &= 
    (\delta r,\: 0)
    =
    (1, 0) \delta r
    \, .
\end{align}
where $D(r,\phi) = dr/d\phi$, then we can get 
\begin{align}
    \cos \psi 
    = 
    \frac{|D|}{\left[D(r,\phi)^{2} + r^{2} B(r)^{-1} \right]^{1/2}}
    \, ,
\end{align}
or more conveniently
\begin{align}
\label{Psi-Approx}
    \tan \psi 
    = 
    \frac{r B(r)^{-1/2}}{|D(r,\phi)|}
    \, .
\end{align}

The one-sided deflection angle is given by $\epsilon_1= \psi-\phi$. 
To obtain the deflection angle, we evaluate it when $\phi = 0$, therefore, the one-sided deflection angle will be $ \epsilon_1 =\psi =\psi_{0}$. 
Noticing the angle $\epsilon_1$ is expected to be very tiny, thus for the LHS of Eq.~\eqref{Psi-Approx}, we have $\tan\psi \simeq \psi$.  
The one-sided deflection angle is: 
\begin{align}
\label{one-sided deflection angle}
    \epsilon_1
    =
    \psi_{0} 
    =
    \frac{r_{\phi=0}}{|D(r_{\phi=0}, 0)|B(r_{\phi=0})^{1/2}}
    \, ,
\end{align}
where the $r_{\phi=0}$ and $\psi_{0}$ are corresponding value of $r$ and $\psi$ when $\phi=0$.
Next, we shall investigate the expression of deflection angle~\eqref{one-sided deflection angle} in our case II.  
We adopt Bodenner and Will's iterative method~\cite{2003Deflection} to solve the nonlinear differential equation~\eqref{differential-equation}.

%%%%%%%%%%%%%%%%%%%%%%%%%%%%%%%%%%%%%%%%%%%%%%%%%%%%%%%%%%%%%%%%%%%%%%%%%%

\subsubsection{Case I with \texorpdfstring{$n=2$}{TEXT}}

According to Eqs.~\eqref{caseII-A}, \eqref{caseII-B} and \eqref{differential-equation}, by setting $n = 2$, we find
\begin{align}
\label{DE-caseI-n2}
    0
    =
    \frac{d^2 u}{d\phi^2} + u 
    - 3 M u^{2}+\frac{16 \alpha M^{3} }{b^{2}} u^{4} 
    - 28 \alpha M^{3} u^{6} 
    \, ,
\end{align}
the expected form of the solution takes $u(\phi) \simeq u_{0}(\phi) + u_{1}(\phi)$, we obtain the following system of differential equations
\begin{align}
\label{DE-system_n2-2}
    0
    &= 
    \frac{d^{2} u_0}{d\phi^{2}} + u_0 
    \, ,
    \\
    0
    &=
    \frac{d^{2} u_1}{d\phi^2} + u_1 - 3 M {u_0}^{2} 
    + \frac{16 \alpha M^{3} }{b^{2}}u_{0}^{4} 
    - 28 \alpha M^{3} u_{0}^{6} 
    \, .
\end{align}
Combining the above two equations, we acquire the solution for $u(\phi)$
\begin{align}
\label{case-I-n2-u}
\begin{split}
    u(\phi) 
    &= 
    \frac{3 M}{2 b^{2}} + \frac{\sin\phi}{b} 
    + \frac{M \cos2\phi}{2 b^{2}} 
    \\
    & \qquad
    + \frac{11 \alpha M^{3}}{4 b^{6}} 
    + \frac{41\alpha M^{3} \cos2\phi}{24 b^{6}} 
    - \frac{13\alpha M^{3} \cos4\phi}{60 b^{6}} 
    + \frac{\alpha M^{3} \cos6\phi}{40 b^{6}}
    \, .
\end{split}
\end{align}
To obtain the deflection angle~\eqref{one-sided deflection angle}, we evaluate the value at $r=r_{\phi=0}$. the Eq.~\eqref{case-I-n2-u} reduce to
\begin{align}
    \frac{1}{r_{\phi=0}} 
    =
    \frac{2 M}{b^{2}} + \frac{64 \alpha M^{3}}{15 b^{6}}
    \, ,
\end{align}
when we take $\phi = 0$. 
Then, we are ready to get the value for $D(r_{\phi=0},0)$:
\begin{align}
\label{n=2,value of D}
\begin{split}
    D(r_{\phi=0},0) 
    &= 
    - r_{\phi=0}^{2} \frac{d u}{d \phi} 
    \bigg|_{\phi = 0} 
    \\
    &= 
    \frac{225 b^{11}}{4 M^{2} \left(15 b^{4} + 32 \alpha M^{2}\right)^{2}}
    \, .
\end{split}
\end{align}
Substituting Eq.~\eqref{n=2,value of D} and the value of $B(r_{0})$ into Eq.~\eqref{one-sided deflection angle}, 
and neglecting the high order of $\alpha$ and $\Lambda$, we get the total deflection angle for $n=2$ in case I:
\begin{align}
\label{light-bending-caseI-n2} 
\begin{split}
    \epsilon ^{\text{I}}_{\text{n=2}} 
    &= 
    2\epsilon_{1} 
    \\
    &\approx 
    \frac{4 M}{b} 
    - \frac{\Lambda b^{3}}{6M}
    + \alpha \cdot \frac{128 M^{3}}{15 b^{5}} 
    \, .
\end{split}
\end{align}

While the first two terms correspond to the deflection angle in GR with the cosmological constant in Eq.~\eqref{light-bending-caseI-n2}, the third term shows the discrepancy with GR occurs at a very high order of $ M/b$. 
Besides, we can compare our result~\eqref{light-bending-caseI-n2} with the similar case in the context of covariant $f(T)$ gravity~\cite{Ren:2021uqb}. 
Except for the contribution of the cosmological constant, 
we can find the same result in the light deflection as Ref.~\cite{Ren:2021uqb}, although we have different geometrical interpretations of gravity in $f(Q)$ and $f(T)$ gravity theories. 
That is, at the level of $(M/b)^{5}$, there are no differences in the light deflection angle between the $f(Q)$ and $f(T)$ gravity.

%%%%%%%%%%%%%%%%%%%%%%%%%%%%%%%%%%%%%%%%%%%%%%%%%%%%%%%%%%%%%%%%%%%%%%%%%%

\subsubsection{Case I with \texorpdfstring{$n = 3$}{TEXT}}

We shall take the same method to consider the case of $n=3$. 
The corresponding solution for Eq.~\eqref{differential-equation} is given by
\begin{align}
\label{case-I-n3-u}
\begin{split}
    u(\phi) 
    &= 
    \frac{3 M}{2 b^{2}} + \frac{\sin \phi}{b}  
    + \frac{M \cos2\phi}{2 b^{2}} 
    \\
    & \qquad
    - \frac{399 \alpha M^{5}}{64 R^{10}} 
    - \frac{259\alpha M^{5} \cos2\phi}{64 b^{10}} 
    + \frac{53\alpha M^{5} \cos4\phi}{80 b^{10}} 
    \\
    & \qquad
    -\frac{717\alpha M^{5} \cos6\phi}{4480 b^{10}} 
    + \frac{13 \alpha M^{5} \cos8\phi}{448 b^{10}} 
    - \frac{\alpha M^{5} \cos10\phi}{384 b^{10}}
     \, .
\end{split}
\end{align}
According to Eq.~\eqref{case-I-n3-u}, we can get the value of $r_{\phi=0}$ by setting $\phi = 0$ 
\begin{align}
    \frac{1}{r_{\phi=0}} 
    = 
    \frac{2 M}{b^{2}} 
    - \frac{1024 \alpha M^{5}}{105 b^{10}}
    \, ,
\end{align}
then $D(r_{\phi=0},0)$ takes the value
\begin{align}
\begin{split}
    D(r_{\phi=0},0) 
    &= 
    - r_{\phi=0}^{2} \frac{d u}{d \phi} \bigg|_{\phi = 0} 
    \\
    &= 
    \frac{11025 b^{19}}{4 M^{2} \left(105 b^{8} 
    - 512 \alpha M^{5}\right)^{2}}
    \, .
\end{split}
\end{align}
Keeping only first-order terms in $\alpha$ and $\Lambda$, we acquire the value of total deflection angle from Eq.~\eqref{one-sided deflection angle}:
\begin{align}
\label{light-bending-caseI-n3} 
    \epsilon^{\text{I}}_{\text{n=3}} 
    \approx 
    \frac{4 M}{b} 
    - \frac{\Lambda b^{3}}{6 M}
    - \alpha \cdot \frac{2048 M^{5}}{105 b^{9}} 
    \, .
\end{align}

\subsubsection{Case II with \texorpdfstring{$n=2, 3$}{TEXT}}

In exactly the same manner as above,
one can calculate the deflection angles for Case II with $n =2,3$, to find the following results:
For case II with $n = 2$,
\begin{align}
\label{n = 2,case II}
    \epsilon^{\text{II}}_{\text{n=2}} 
    \approx 
    \frac{4 M}{b} 
    - \frac{\Lambda b^{3}}{6 M}
    + \alpha \cdot \frac{40M}{ b^{3}} 
    \, .
\end{align}
For case II with $n = 3$,
\begin{align}
\label{n = 3,case II}
    \epsilon^{\text{II}}_{\text{n=3}} 
    \approx 
    \frac{4 M}{b} 
    - \frac{\Lambda b^{3}}{6 M}
    - \alpha \cdot \frac{560 M}{3 b^{5}} 
    \, .
\end{align}

Comparing Eqs.~\eqref{light-bending-caseI-n2} and \eqref{light-bending-caseI-n3} with Eqs.~\eqref{n = 2,case II} and \eqref{n = 3,case II},
we find a large discrepancy in light deflection angles in two different choices of affine connections. For the same value of $n$,
Case I predicts the higher order corrections of $M/b$ to the deflection angle,
and thus, the deviations from GR prediction in Case I is generally less than those in Case II.

%%%%%%%%%%%%%%%%%%%%%%%%%%%%%%%%%%%%%%%%%%%%%%%%%%%%%%%%%%%%%%%%%%%%%%%%%%

\subsection{Shapiro time delay}
\label{sec:Shapiro Delay}

In this subsection, we study the Shapiro effect~\cite{Shapiro:1964uw}.
The motion of a photon is retarded in the presence of gravity, and it predicts the observables time delay in signals.
We consider a radar signal emitted from the Earth, swept across the Sun, captured by a spacecraft and then returned to Earth. 
According to Eqs.~\eqref{light-bending-1st-order-DE}-\eqref{energy-momentum-null}, the trajectory of the signal can be reduce to:
\begin{align}
\label{time-delay-eq}
    \frac{ d t}{d r} 
    = 
    \left[\frac{A(r)}{B(r)} 
    \left( 1 - \frac{A(r) b^{2}}{r^{2}} \right) \right]^{-1/2}
    \, ,
\end{align}
where $b = L/E$ is the impact parameter. 
At the point $r=r_{0}$ for the closest approach to the Sun, we have $dr/dt = 0$ which leads to the condition
\begin{align}
    b^{2} = \frac{r_{0}^{2}}{A(r_{0})}
    \, .
\end{align}
Then, the time required for a signal propagating from $r$ to $r_{0}$ is given as
\begin{align}
\label{time integral}
    t(r,r_{0}) 
    &= 
    \int_{r_{0}}^{r} \left[\frac{A(r)}{B(r)} \left(1 - \frac{A(r) r_{0}^{2}}{A(r_{0}) r^{2} } \right) \right]^{-1/2} \: d r 
    \, .
\end{align}

The time delay is maximized when the reflector is in superior conjunction and the signal passes very close to the Sun 
($r_{0} \approx r_{\odot}$, where $r_{\odot} $ is the radius of Sun).
In this case, the maximum time delay for the round-trip travel time is given by:
\begin{align}
\label{round-trip time}
    \Delta t 
    = 
    2 
    \left[
        t(r_{e}, r_{\odot}) +  t(r_{s}, r_{\odot}) 
        - \sqrt{r_{e}^{2} - r_{\odot}^{2}} 
        - \sqrt{r_{s}^{2} - r_{\odot}^{2}}
    \right]
    \, ,
\end{align}
where $r_{e}$ and $r_{s}$ denote the distance from Earth and spacecraft to the Sun, respectively.
By making use of the expression~\eqref{time integral} and the facts that $r_{e} \gg r_{\odot}$ and $r_{s} \gg r_{\odot}$, Eq.~\eqref{round-trip time} leads to
\begin{align}
\label{time delay}
\begin{split}
    \Delta t 
    &\simeq 
    4M + 4M \ln \left( \frac{4 r_{e} r_{s}}{r_{\odot}^{2}} \right) 
    + \frac{2 \Lambda}{9} (r_{e}^{3} + r_{s}^{3}) 
    \\
    & \qquad
    + \frac{2 \Lambda M}{3} (2 r_{e}^{2} + 2 r_{s}^{2} + r_{\odot}^{2}) 
    % \\
    % & \qquad
    + 2\left[ T_{n}(r_{e}, r_{\odot}) + T_{n}(r_{s}, r_{\odot})\right]
    \, .
\end{split}
\end{align}
We have used Robertson expansion in the above calculation and also kept the term $\mathcal{O}(\Lambda M)$ to calculate the fractional frequency shift in the next section.  

$T_{n}(r, r_{\odot})$ in Eq.~\eqref{time delay} denotes the corresponding terms in each of our cases as follows.
For case I with $n=2$, we have 
\begin{align}
    T_{2}(r, r_{\odot}) 
    \simeq 
    \alpha \sqrt{\frac{r-r_{\odot}}{r+r_{\odot}}} 
    \left( \frac{16M^{3}}{3r_{\odot}^{4}} + \frac{28M^{3}}{15 r^{3} r_{\odot}} + \frac{28 M^{3}}{r^{2} r_{\odot}^{2}} + \frac{68M^{3}}{15 r r_{\odot}^{3}}\right)
    \, .
\end{align}
Using the conditions $r_{e} \gg r_{\odot}$ and $r_{s} \gg r_{\odot}$, one can then find that in this case, Eq.~\eqref{time delay} takes the form
\begin{align}
\label{Shapiro,case I,n=2}
    \Delta t^{\text{I}}_{\text{n=2}} 
    &\approx 
    4M + 4M \ln \left( \frac{4 r_{e} r_{s}}{r_{\odot}^{2}} \right) 
    + \frac{2 \Lambda}{9} (r_{e}^{3} + r_{s}^{3}) 
    \\
    & \qquad
    + \frac{2 \Lambda M}{3} (2 r_{e}^{2} + 2 r_{s}^{2} + r_{\odot}^{2}) 
    + \alpha \cdot \frac{64 M^{3}}{3 r_{\odot}^{4}}
    \, .
\end{align} 
For case I with $n=3$, we have
\begin{align}
\begin{split}
    T_3\left(r, r_{\odot}\right) 
    & \simeq 
    -\alpha \sqrt{\frac{r-r_{\odot}}{r+r_{\odot}}}
    \left[
        \frac{384 M^5}{35 r_{\odot}^8}  \left(1+\frac{288 r_{\odot}}{253 r}\right)+\frac{44 M^5}{21 r^6 r_{\odot}^2}\left(1+\frac{r_{\odot}}{r}\right)
    \right. 
    \\
    & \qquad \qquad \qquad \qquad
    \left.
        + \frac{292 M^5}{105 r^4 r_{\odot}^4}\left(1+\frac{r_{\odot}}{r}\right)
        +\frac{436 M^5}{105 r^2 r_{\odot}^6}\left(1+\frac{r_{\odot}}{r}\right)
    \right]
    \, .
\end{split}
\end{align} 
The corresponding retardation of light is given as
\begin{align}
\label{Shapiro,case I,n=3}
\begin{split}
    \Delta t^{\text{I}}_{\text{n=3}}
    &\approx 
    4M + 4M \ln \left( \frac{4 r_{e} r_{s}}{r_{\odot}^{2}} \right) 
    \\
    &
    + \frac{2 \Lambda}{9} (r_{e}^{3} + r_{s}^{3}) 
    + \frac{2 \Lambda M}{3} (2 r_{e}^{2} + 2 r_{s}^{2} + r_{\odot}^{2}) 
    - \alpha \cdot \frac{1536 M^{5}}{35 r_{\odot}^{8}} 
    \, .
\end{split}
\end{align}

Using the same method as above, we find for case II with $n=2$,
\begin{align}
\label{Shapiro,case II,n=2}
\begin{split}
    \Delta t^{\text{II}}_{\text{n=2}}
    &\approx 
    4M + 4M \ln \left( \frac{4 r_{e} r_{s}}{r_{\odot}^{2}} \right) 
    \\
    &
    + \frac{2 \Lambda}{9} (r_{e}^{3} + r_{s}^{3}) 
    + \frac{2 \Lambda M}{3} (2 r_{e}^{2} + 2 r_{s}^{2} + r_{\odot}^{2}) 
    + \alpha \cdot \frac{160 \pi}{r_{\odot}}
    \, ,
\end{split}
\end{align}
and for case II with $n=3$,
\begin{align}
\label{Shapiro,case II,n=3}
\begin{split}
    \Delta t^{\text{II}}_{\text{n=3}}
    &\approx 
    4M + 4M \ln \left( \frac{4 r_{e} r_{s}}{r_{\odot}^{2}} \right) 
    \\
    &
    + \frac{2 \Lambda}{9} (r_{e}^{3} + r_{s}^{3}) 
    + \frac{2 \Lambda M}{3} (2 r_{e}^{2} + 2 r_{s}^{2} + r_{\odot}^{2}) 
    + \alpha \cdot \frac{1792 \pi}{3 r_{\odot}^{3}}
    \, .
\end{split}
\end{align}

%%%%%%%%%%%%%%%%%%%%%%%%%%%%%%%%%%%%%%%%%%%%%%%%%%%%%%%%%%%%%%%%%%%%%%%%%%

\subsection{Cassini constraint}
\label{sec:Cassini Experiment}

In relation to the Shapiro time delay, the Cassini spacecraft constrained the relative change in frequency~\cite{bertotti2003test}.
The corresponding two-way fractional frequency shift is given by the following expression:
\begin{align}
    y = \frac{\Delta \nu}{\nu} = \frac{d \Delta t}{dt}~,
\end{align}
where $\Delta t$ is the retardation of signal found in the Subs.~\ref{sec:Shapiro Delay}. 
The GR contribution to the fractional frequency shift is
\begin{align}
    y_{\text{GR}}& = -\frac{8 M}{b} v_{e}
    \, ,
\end{align}
here, $v_{e}\approx b'(t)$ denotes the orbital velocity of the Earth, and $b(t)$ is the impact parameter of the signal to the Sun. 

Following the results in the previous subsection, we calculate fractional frequency shift $y$ for each of our models:
For Case I with $n=2$,
\begin{align}
\label{Cassini,case I,n=2}
    y^{\text{I}}_{\text{n=2}}
    \approx 
    -\frac{8 M}{b}v_{e} 
    + \frac{4\Lambda M}{3} b v_{e} 
    - \alpha \cdot \frac{256 M^{3}}{b^{5}} v_{e}
    \, ;
\end{align}
For Case I with $n=3$,
\begin{align}
\label{Cassini,case I,n=3}
    y^{\text{I}}_{\text{n=3}}
    \approx 
    -\frac{8 M}{b} v_{e} 
    + \frac{4\Lambda M}{3} b v_{e} 
    + \alpha \cdot \frac{12288 M^{5}}{b^{9}} v_{e}
    \, ;
\end{align}
For case II with $n=2$,
\begin{align}
\label{Cassini,case II,n=2}
    y^{\text{II}}_{\text{n=2}}
    \approx
    - \frac{8 M}{b} v_{e} 
    + \frac{4\Lambda M}{3} b v_{e} 
    - \alpha \cdot \frac{160 \pi}{b^{2}} v_{e}
    \, ;
\end{align}
For case II with $n=3$,
\begin{align}
\label{Cassini,case II,n=3}
    y^{\text{II}}_{\text{n=3}}
    \approx 
    -\frac{8 M}{b} v_{e} 
    + \frac{4\Lambda M}{3} b v_{e} 
    - \alpha \cdot \frac{1792 \pi}{b^{4}} v_{e}
    \, .
\end{align}
We note that the frequency shift takes peak value for the time of solar conjunction $t_0$, when the spacecraft, the Sun, and the Earth are almost aligned. 
The impact parameter at this point is approximately equal to the solar radii $b(t_0)\approx r_{\odot}$.

%%%%%%%%%%%%%%%%%%%%%%%%%%%%%%%%%%%%%%%%%%%%%%%%%%%%%%%%%%%%%%%%%%%%%%%%%%

\subsection{Gravitational redshift}
\label{sec:Gravitational Redshift}

At last, we consider the gravitational redshift effect. 
Considering a photon propagates from the position $r_{1}$ to $r_{2}$ ($r_{1} < r_{2}$), 
the gravitational redshift $z$ is expressed as
\begin{align}
    z 
    \equiv
    \frac{\nu_{2}}{\nu_{1}} - 1 = 
    \frac{A(r_2)^{1/2}}{A(r_1)^{1/2}} - 1
    \, ,
\end{align}
where $\nu_{1}$ and $\nu_{2}$ are the frequencies of photon observed at $r_{1}$ and $r_{2}$ respectively.
Then, in the spherically symmetric spacetime given by Eq.~\eqref{general case}, the redshift can be written as
\begin{align}
    z 
    \approx 
    \frac{a_{1}(r_{1}) M }{2 r_{1}} 
    - \frac{a_{1}(r_{2}) M }{2 r_{2}}
    - \frac{a_{2}(r_{1}) M^{2} }{2 r_{1}^{2}}
    + \frac{a_{2}(r_{2}) M^{2} }{2 r_{2}^{2}}
    \, .
\end{align}

For case I, the result of gravitational redshift is given as
\begin{align}
\label{redshift,case I}
\begin{split}
    z^{\text{I}} 
    & \approx 
    \frac{M}{{r_1}}
    - \frac{M}{{r_2}} 
    + \frac{1}{6}\Lambda \left({r_1}^2 - {r_2}^2\right) 
    \\
    & \qquad
    - \alpha 
    \cdot
    \frac{(-1)^{n} 2^{n - 1} n}{4n - 3} \cdot
    M^{2n - 1} \left(r_{1}^{3 - 4n} - r_{2}^{3-4n}\right)
    \, .
\end{split}
\end{align}
For case II, the gravitational redshift takes the form
\begin{align}
\label{redshift,case II}
\begin{split}
    z^{\text{II}} 
    & \approx 
    \frac{M}{{r_1}}
    - \frac{M}{{r_2}} 
    + \frac{1}{6}\Lambda \left({r_1}^2 - {r_2}^2\right) 
    \\
    & \qquad
    - \alpha \cdot \frac{2^{3n - 2}}{3-2n} \cdot 
    \left(r_{1}^{2-2n} - r_{2} ^{2 - 2n}\right)
    \, ,
\end{split}
\end{align}

%%%%%%%%%%%%%%%%%%%%%%%%%%%%%%%%%%%%%%%%%%%%%%%%%%%%%%%%%%%%%%%%%%%%%%%%%%
%%%%%%%%%%%%%%%%%%%%%%%%%%%%%%%%%%%%%%%%%%%%%%%%%%%%%%%%%%%%%%%%%%%%%%%%%%

\section{Solar System constraints}
\label{solar system constraints}

In the previous section, we have established formulations to investigate the Solar System tests in $f(Q)$ gravity theory.
We have considered the perihelion precession, light deflection, Shapiro time delay, Cassini constraint and gravitational redshift, 
and they show tiny but nonzero deviations from predictions in GR.
In this section, we will investigate the constraint on the parameter $\alpha$ by using the latest observations results in the Solar System. 
For our analysis, We choose $\Lambda \simeq  10^{-46} $ $\text{km}^{-2}$ and $ M_{\odot} \simeq  1.4766 $ $\text{km}$, $r_{\odot} \simeq 6.957 \times 10^{5}$ $\text{km}$ for the Sun, while $M \simeq  4.4284 \times 10^{-6}$ $\text{km}$ for the Earth.

%%%%%%%%%%%%%%%%%%%%%%%%%%%%%%%%%%%%%%%%%%%%%%%%%%%%%%%%%%%%%%%%%%%%%%%%%%

\subsection{Perihelion precession}

We compare the predicted precession angles in Eqs.~\eqref{case I-solution} and \eqref{case II-solution} with the observed residual precession angles for Mercury, Venus, and Earth shown in Table~\ref{Perihelion-precession-data}~\footnote{The observational data are available in the current planetary ephemerides EPM2011~\cite{Pitjeva:2013fja,2010IAUS..261..170P}.}
The residual precession angle in Table~\ref{Perihelion-precession-data} refers to the precession angle that excludes the known Newtonian effects, such as the gravitational tugs.
Accordingly, we derive constraints of $\alpha$ for each planet, as we show in Table~\ref{upper limit alpha}.

\begin{table}[htbp]
\centering\setlength{\tabcolsep}{8pt}
\begin{tabular}{c c c c c c}
\hline\hline
\rule{0pt}{3ex} Planet &  $a$ (au) &  Eccentricity &  Rev./cty. & $\Delta\phi_{\text{GR}}$($^{\prime\prime}$/per rev.)& Corr. ($^{\prime\prime}$/cty.)    \\ [0ex] \hline
\rule{0pt}{3ex} 
  Mercury   & 0.3871 & 0.206 & 414.9378 & 0.1034 & -0.0040 $\pm $ 0.0050  \\
 Venus   & 0.7233 & 0.007 & 162.6016 & 0.0530  & 0.0240 $\pm $ 0.0330 \\
 Earth  & 1.0000  & 0.017  & 100.0000 & 0.0383  &  0.0060 $\pm $ 0.0070 \\
\hline\hline
\end{tabular}
\caption{
    Perihelion precession data of planets in the Solar System.
    Table headers from left to right are semi-major axis $a$ (astronomical unit), eccentricity $e$, revolutions (century), residual precession angle (arcsec per revolution) predicted by the general relativistic term, and the observed correction for the residual precession angle of the three solar planets respectively.
    }
\label{Perihelion-precession-data}
\end{table}

\begin{table}[htbp]
\resizebox{\textwidth}{!}{%
\begin{tabular}{cl|c|c|c}
\hline\hline
\multicolumn{2}{c|}{} & Mercury & Venus & Earth \\ \hline
\multicolumn{1}{c|}{\multirow{2}{*}{Case I}} & $n=2$ & $-9.07\times 10^{18}<\alpha < 1.01\times 10^{18}$        &  $-3.36\times 10^{20}<\alpha < 2.13\times 10^{21}$     &   $-2.21\times 10^{20}<\alpha <2.88\times 10^{21}$    
\\ 
\multicolumn{1}{c|}{}                      & $n=3$ &  $-7.29\times 10^{47}<\alpha < 6.56\times 10^{48}$       &  $-2.23\times 10^{52}< \alpha < 3.52 \times 10^{51}$      &  $-1.10\times 10^{53}<\alpha < 8.46\times 10^{51}$  
\\ \hline
\multicolumn{1}{l|}{\multirow{2}{*}{Case II}} & $n=2$ & $-8.56\times 10^{-5}<\alpha <9.52\times 10^{-6}$ & $-4.26\times 10^{-4}<\alpha <2.70\times 10^{-3}$     & $-1.06\times 10^{-4}<\alpha <1.38\times 10^{-3}$      
\\ 
\multicolumn{1}{l|}{} & $n=3$ & $-1.65\times 10^{10}<\alpha <1.83\times 10^{9}$ & $-3.12\times 10^{11}<\alpha <1.98\times 10^{12}$    &  $-1.49\times 10^{11}<\alpha <1.94\times 10^{12}$     
\\ \hline\hline
\end{tabular}%
}
\caption{
    Constraint on $\alpha$ from the perihelion precessions of Mercury, Venus, and Earth.
}
\label{upper limit alpha}
\end{table}

By comparing the values in Table~\ref{upper limit alpha}, 
the perihelion precession of Mercury gives the stringent constraint on $\alpha$,
and we can summarize the constraint on $\alpha$ as follows: 
For case I, $n =2$, 
\begin{align}
   -9.07 \times 10^{18} \:\text{km}^{2}< \alpha < 1.01 \times 10^{18} \:\text{km}^{2}~,
\end{align}
 whilst for $n = 3$
\begin{align}
    -7.29 \times 10^{47}\:\text{km}^{4} < \alpha < 6.56 \times 10^{48} \:\text{km}^{4}~.
\end{align}

For case II, $n = 2$, 
\begin{align}
\label{Perihelion precession case I n = 2}
    -8.56 \times 10^{-5} \:\text{km}^{2}< \alpha < 9.52 \times 10^{-6} \:\text{km}^{2}~,
\end{align}
 whilst for $n = 3$
\begin{align}
    -1.65 \times 10^{10} \:\text{km}^{4}< \alpha < 1.83 \times 10^{9} \:\text{km}^{4}~.
\end{align}

One may notice that the upper bound of $|\alpha|$ for $n=3$ looks extremely large.
However, $\alpha$ is not a dimensionless quantity that can be made arbitrarily large or small by a simple change of units.
Case I deviates less from GR compared to Case II, and therefore, we have weaker constraints on Case I than those on Case II.

%%%%%%%%%%%%%%%%%%%%%%%%%%%%%%%%%%%%%%%%%%%%%%%%%%%%%%%%%%%%%%%%%%%%%%%%%%

\subsection{Light deflection}

We can utilize observational values of the light deflection from the Sun. 
The light deflation is related to the PPN parameter $\gamma$,
\begin{align}
    \epsilon 
    =
    \gamma \frac{4M}{b}
    \, .
\end{align}
We set $b \approx r_{\odot} $ as the solar radius, and we utilize the observational constraint $\gamma =0.99992 \pm 0.00012$~\cite{Lambert:2009xy}. 

For Case I with $n = 2$,
Eq.~\eqref{light-bending-caseI-n2} gives the constraint on $\alpha$ as
\begin{align}
    -1.01\times 10^{19}\:\text{km}^{2} < \alpha <2.22 \times 10^{18} \:\text{km}^{2}
    \, ,
\end{align}
and with $n = 3$, Eq.~\eqref{light-bending-caseI-n3} gives
\begin{align}
    -9.47 \times 10^{40} \:\text{km}^{4}< \alpha < 4.73\times 10^{41} \:\text{km}^{4}
    \, .
\end{align}
For Case II with $n = 2$, in the same way as in Case I, we find
\begin{align}
    -9.68 \times 10^{6}\:\text{km}^{2} < \alpha < 1.94 \times 10^{6} \:\text{km}^{2}
    \, ,
\end{align}
and with $n = 3$,
\begin{align}
    -2.01 \times 10^{17}\:\text{km}^{4} < \alpha < 1.00 \times 10^{18} \:\text{km}^{4}
    \, .
\end{align}

%%%%%%%%%%%%%%%%%%%%%%%%%%%%%%%%%%%%%%%%%%%%%%%%%%%%%%%%%%%%%%%%%%%%%%%%%%

\subsection{Shapiro time delay}

We use the observations from the Viking mission on Mars, the radio signal is emitted from Earth to Mars and back, the signal passes close to the Sun's surface $b \approx r_{\odot}$, we know that~\cite{reasenberg1979viking}
\begin{align}
    \frac{\Delta t_{obs}}{\Delta t_{GR}} = 1.000 \pm 0.001
    \, .
\end{align}
The emitter and reflector distances to the Sun are calculated from the semi-major axis $a$ (au), eccentricity $e$ of the Earth and Mars in Table~\ref{Perihelion-precession-data}, respectively. 
Then, using the expression in our model Eq.~\eqref{Shapiro,case II,n=2}-\eqref{Shapiro,case I,n=3}, one could get the constraints for $\alpha$:

For case I, $n = 2$
\begin{align}
    -2.73 \times 10^{20} \:\text{km}^{2}< \alpha < 2.73 \times 10^{20} \:\text{km}^{2}
    \, ,
\end{align}
for $n = 3$, we get
\begin{align}
    -1.42 \times 10^{43} \:\text{km}^{4}< \alpha < 1.42 \times 10^{43} \:\text{km}^{4}
    \, .
\end{align}

For case II $n=2$, we have
\begin{align}
    -1.11 \times 10^{2} \:\text{km}^{2}< \alpha < 1.11 \times 10^{2} \:\text{km}^{2}
    \, ,
\end{align}
for $n = 3$
\begin{align}
    -1.43 \times 10^{13}\:\text{km}^{4} < \alpha < 1.43 \times 10^{13} \:\text{km}^{4}
    \, .
\end{align}

%%%%%%%%%%%%%%%%%%%%%%%%%%%%%%%%%%%%%%%%%%%%%%%%%%%%%%%%%%%%%%%%%%%%%%%%%%

\subsection{Cassini constraint}

The Cassini spacecraft gives the constraint of the PPN parameter $\gamma$ as $\gamma=1 + (2.1 \pm 2.3)\times 10^{-5}$.
The signal corresponding fractional frequency shift is $ y_{\text{Obs}}\simeq y_{\text{GR}} \pm 10^{-14}$, where $y_{\text{GR}}=10^{-10}$~\cite{bertotti2003test}. 
We consider $v_{e} \simeq 30 $ km/s for the Earth’s velocity, the impact parameter $b \simeq r_{\odot}$ and the speed of light $c \simeq 3 \times 10^{8}$ m/s. 
Inputting these values into Eqs.~\eqref{Cassini,case I,n=2} and \eqref{Cassini,case II,n=3}, we obtain the following constraints on $\alpha$.

For Case I with $n = 2$,
\begin{align}
    -1.98 \times 10^{19}\:\text{km}^{2} < \alpha < 1.98 \times 10^{19} \:\text{km}^{2}
    \, ,
\end{align}
and for Case I with $n = 3$,
\begin{align}
    -4.43 \times 10^{40} \:\text{km}^{4}< \alpha < 4.43 \times 10^{40} \:\text{km}^{4}
    \, .
\end{align}
For Case II with $n=2$,
\begin{align}
% \label{\frac case II n = 2}
    -9.63 \times 10^{1}\:\text{km}^{2} < \alpha < 9.63 \times 10^{1} \:\text{km}^{2}
    \, ,
\end{align}
for Case Ii with $n = 3$,
\begin{align}
    -4.16 \times 10^{12}\:\text{km}^{4} < \alpha < 4.16 \times 10^{12} \:\text{km}^{4}
    \, .
\end{align}

%%%%%%%%%%%%%%%%%%%%%%%%%%%%%%%%%%%%%%%%%%%%%%%%%%%%%%%%%%%%%%%%%%%%%%%%%%

\subsection{Gravitational redshift}

To constrain the $\alpha$ by the gravitational redshift, 
we utilize the experiment performed by Pound and Rebka, which determined the redshift due to the Earth on a particle falling a height of 22.5 m.
Measured redshift is $z = (2.57 \pm 0.26) \times 10^{-15}$~\cite{PhysRevLett.3.439}. 
We input $z$ and $r_{1} \simeq 6.3570 \times 10^{3} \text{~km}$ in Eqs.~\eqref{redshift,case I} and \eqref{redshift,case II} and compute the constraint on $\alpha$.

For Case I with $n = 2$, 
\begin{align}
    -3.08 \times 10^{24}\:\text{km}^{2} < \alpha < 1.31 \times 10^{24} \:\text{km}^{2}
    \, ,
\end{align}
and for Case I with $n = 3$, 
\begin{align}
    -3.65 \times 10^{49}\:\text{km}^{4} < \alpha < 8.54 \times 10^{49} \:\text{km}^{4}
    \, .
\end{align}
For case II with $n = 2$, 
\begin{align}
\label{Gravitational redshift case II n = 2}
    -5.55 \times 10^{-5}\:\text{km}^{2} < \alpha < 1.30 \times 10^{-4} \:\text{km}^{2}
    \, ,
\end{align}
and for Case II with $n = 3$,
\begin{align}
\label{Gravitational redshift case II n = 3}
    -4.21 \times 10^{2}\:\text{km}^{4} < \alpha < 9.85 \times 10^{2} \:\text{km}^{4}
    \, .
\end{align}

%%%%%%%%%%%%%%%%%%%%%%%%%%%%%%%%%%%%%%%%%%%%%%%%%%%%%%%%%%%%%%%%%%%%%%%%%%
%%%%%%%%%%%%%%%%%%%%%%%%%%%%%%%%%%%%%%%%%%%%%%%%%%%%%%%%%%%%%%%%%%%%%%%%%%

\section{Conclusions}
\label{conclusions}

$f(Q)$ gravity brings new life to GR by providing a novel interpretation of gravity from a different geometrical perspective. 
By considering the framework of different affine connections in $f(Q)$ gravity, interesting physical effects beyond GR may arise, which would provide new avenues of search for model proposals in $f(Q)$ gravity.
In this work, we investigate several effects of $f(Q)$ gravity on the astronomical observation and experiments conducted in the Solar System. 
We focus on the model in the form $f(Q) = Q + \alpha Q^{n} -2\Lambda$, where the parameter $\alpha$ embodies the deviation from GR, and the cosmological constant $\Lambda$ is included for completeness. 

First, applying the perturbative and weak-gravity approximation, we have derived the static and spherically symmetric solutions~\eqref{caseI-A}-\eqref{caseII-B} in our model for two different cases of affine connections. 
The outcome of spherical solutions shows that the deviation of the metric solution from the Schwarzschild-de Sitter turned out to be in the power of radial coordinate: $r^{3-4n}$ in Case I and $r^{2-2n}$ in Case II.
We have confirmed that for $n=2,3$, these deviations decrease as the radial coordinate $r$ increases.

Next, using the obtained solutions, we have formulated paths for massive and massless particles to calculate the perihelion precession and light deflection in the Subs.~\ref{sec:Perihelion Precession} and~\ref{sec:light-bending}. 
We have found that for Case II with $n = 2$, the deviation from GR with respect to the light deflection is given at the order of $(M/b)^3$ in Eq.~\eqref{n = 2,case II}.
The deflection angle in Case I with $n = 2$ deviates from GR prediction at the order of $(M/b)^5$, which is consistent results in Ref.~\cite{Ren:2021uqb}, 
suggests that Case I deviates less from GR compared to those in Case II.

Finally, to obtain further constraints on $\alpha$,
we have developed our calculations in perihelion precession and light deflection and applied them to other experiments:
The calculation of Shapiro delay is presented in Subs.~\ref{sec:Shapiro Delay}; Cassini constraint in Subs.~\ref{sec:Cassini Experiment}; and the gravitational redshift in Subs.\ref{sec:Gravitational Redshift}. 
We have explored the deviations from GR predictions for each type of experiment. 
Our results in Eqs.~\eqref{Shapiro,case I,n=3}-\eqref{Shapiro,case II,n=3}, 
Eq.~\eqref{Cassini,case I,n=2}-\eqref{Cassini,case II,n=3}, 
Eq.~\eqref{redshift,case I}-\eqref{redshift,case II} 
show that the different contributions from $\alpha Q^n$ emerge depending on the choice of the affine connection; for instance, different dependence on the impact parameter $b$ and source mass $M$.
Nonetheless, concerning the magnitude of $\Lambda$ is very tiny, it is possible to constrain the $\alpha$ through experimental data.

In Sec.~\ref{solar system constraints}, we used the latest observational data to constrain the $\alpha$ and obtained a range of values that satisfies each observation data set. 
As we can see, for case II, $n = 2,3$, the Gravitational redshift~\eqref{Gravitational redshift case II n = 2}-\eqref{Gravitational redshift case II n = 3} give the best restrictions on $\alpha$, which the accuracy takes the order of magnitude $10^{-5}-10^{-4}\text{~km}^{2}$ for $n=2$ and $10^{2}\text{~km}^{4}$ for $n=3$. 
This is partially because the other tests are performed on a relatively large scale in the Solar System, yielding a negligible higher order term ${\cal O}(1/r^{n})$. 
However, the gravitational redshift test is conducted in the vicinity of the Earth, thus avoiding this issue.

It is worth mentioning that for Case I with $n=3$, since the correction term is in power form $\sim Q^{3}$, we have found loose constraints on $\alpha$; for all experiments we considered, the upper bound of $\alpha$ turns out to be enormous. 
As we mentioned before, the higher order of $Q$ implies that the contribution from this factor is extremely small. 
Although there is a significant difference in magnitude between the values of $\alpha$ for the two choices of $n$, it is mandatory to check the linear term of the nonmetricity scalar $Q$ still dominant in the Lagrangian of our model $Q \gg \alpha Q^{n}$. 
As an example, for case I, we consider the case of the perihelion precession of Mercury, utilize the obtained constraint
~\eqref{constants-caseI-n2} -~\eqref{constants-caseI-n3} in the nonmetricity scalar~\eqref{nonmetricity scalar-caseI}, we get $|Q| \sim 4.61 \times 10^{-31} \text{km}^{-2} $ which is much bigger than $|\alpha Q^{2}| \sim 2.15 \times 10^{-43} \text{km}^{-2} $ and $|\alpha Q^{3}| \sim 4.35 \times 10^{-51} \text{km}^{-2}$.
For case II, we use the obtained constraint~\eqref{constants-caseII-n2} -~\eqref{constants-caseII-n3} and nonmetricity scalar~\eqref{nonmetricity scalar-caseII} can give us $|Q| \sim 2.60 \times 10^{-15} \text{km}^{-2} $ which is much bigger than $|\alpha Q^{2}| \sim 6.44 \times 10^{-35} \text{km}^{-2} $ and $|\alpha Q^{3}| \sim 1.73 \times 10^{-41} \text{km}^{-2} $.
The above can also be directly checked in the spherically symmetric solutions in Eqs.~\eqref{caseI-A}-\eqref{nonmetricity scalar-caseII}.
When we take a relatively larger value for $n$, the correction terms in these expressions decrease by the power law.

We conclude the Solar-System constraints on the model $f(Q) = Q + \alpha Q^n - 2\Lambda$. 
Gathering all the constraints shown in the Sec.~\ref{solar system constraints}, 
we find that the parameter $\alpha$ should satisfy the following constraints:
In Case I,
\begin{align}
\label{constants-caseI-n2}
\begin{split}
    -9.07 \times 10^{18}\:\text{km}^{2} < \alpha < 1.01 \times 10^{18} \:\text{km}^{2}
    \quad (n=2)
    \, ,
\end{split}
    \\
\label{constants-caseI-n3}
\begin{split}
    -4.43 \times 10^{40} \:\text{km}^{4}< \alpha < 4.43\times 10^{40} \:\text{km}^{4}
    \quad (n=3)
    \, ;
\end{split}
\end{align}
In Case II,
\begin{align}
\label{constants-caseII-n2}
\begin{split}
    -5.55 \times 10^{-5} \:\text{km}^{2}< \alpha < 9.52 \times 10^{-6} \:\text{km}^{2}
    \quad (n=2)
    \, ,
\end{split}
\\
\label{constants-caseII-n3}
\begin{split}
    -4.21 \times 10^{2}\:\text{km}^{4} < \alpha < 9.85 \times 10^{2} \:\text{km}^{4}
    \quad (n=3)
    \, .
\end{split}
\end{align}

It should be pointed out that, technically speaking, any Lagrangian of $f(Q)$ gravity can be power-series expanded as $f(Q)\sim Q+Q^2+Q^3+\cdots $.
In future works, we can extend the current analysis by considering more general cases of nometricity corrections of the Lagrangian or including more possible cases of admissible connections.

%%%%%%%%%%%%%%%%%%%%%%%%%%%%%%%
%%%%%%%%%%%%%%%%%%%%%%%%%%%%%%%

\acknowledgments
The authors thank Taotao Qiu for helpful discussions. 
T.K. is supported by the National Key R\&D Program of China (No.~2021YFA0718500) and Grant-in-Aid of Hubei Province Natural Science Foundation (No.~2022CFB817).

%%%%%%%%%%%%%%%%%%%%%%%%%%%%%%%
%%%%%%%%%%%%%%%%%%%%%%%%%%%%%%%

\bibliographystyle{apsrev4-1}
\bibliography{References}

\end{document}